\documentclass[aps,prb,twocolumn]{revtex4}
\usepackage{bm}
\usepackage{graphicx}

\begin{document}

\title{Weak localization in arrays of metallic quantum dots}
\author{Dmitri S. Golubev and
Andrei D. Zaikin}
\affiliation{Forschungszentrum Karlsruhe, Institut f\"ur Nanotechnologie,
76021, Karlsruhe, Germany\\
DFG-Center for Functional Nanostructures (CFN),
Universit\"at Karlsruhe, 76128 Karlsruhe, Germany\\
I.E. Tamm Department of Theoretical Physics, P.N.
Lebedev Physics Institute, 119991 Moscow, Russia}

\begin{abstract}

Combining scattering matrix formalism with non-linear $\sigma$-model 
technique we analyze weak localization effects in arrays of 
chaotic quantum dots connected via barriers
with arbitrary distribution of channel transmissions. 
With the aid of our approach we evaluate magnetoconductance
of two arbitrarily connected quantum dots 
as well as of $N\times M$ arrays of identical quantum dots. 

\end{abstract}

\maketitle

\section{Introduction}

Quantum interference of electrons is fundamentally
important for electron transport in disordered conductors \cite{Berg,AA,CS}.
Quantum coherent effects are mostly pronounced
at low temperatures in which case certain interaction mechanisms are
effectively ``frozen out'' and, hence, cannot anymore restrict the ability of 
electrons to interfere. At the same time, there exists at least one mechanism,
electron-electron interactions, which remains important down to lowest
temperatures and may destroy quantum interference of electrons down to $T=0$.
It is, therefore, highly desirable to formulate a general theoretical
formalism which would allow to describe electron interference effects in
the presence of disorder and electron-electron interactions at any
temperature, including the limit $T \to 0$.

In a series of papers \cite{GZ} we offered such an approach
which extends Chakravarty-Schmid description \cite{CS} of weak
localization (WL) and generalizes Feynman-Vernon path integral
influence functional technique to fermionic systems with disorder
and interactions. With the aid of this technique it turned out to
be possible to quantitatively explain low temperature saturation
of WL correction to conductance $\delta G^{WL} (T)$ commonly
observed in diffusive metallic wires \cite{MohGre}. It was
demonstrated \cite{GZ} that this saturation effect is caused
by electron-electron interactions.

It is worth pointing out that low temperature saturation of WL
correction and of the electron decoherence time $\tau_{\varphi}$
(extracted from $\delta G^{WL} (T)$ or by other means) has been
repeatedly observed not only in metallic wires but also in
virtually any type of disordered conductors ranging from
individual quantum dots \cite{dephdots} to very strongly
disordered 3d structures and granular metals \cite{BL}. It is
quite likely that in all these systems we are dealing with {\it
the same} fundamental effect of electron-electron interactions.
In order to support (or discard) this conjecture it is
necessary to develop a unified theoretical description which would
cover essentially all types of disordered conductors. Although the
approach \cite{GZ} is formally an exact procedure treating
electron dynamics in the presence of disorder and interactions, in
some cases, e.g., for quantum dots and granular metals, it can be
rather difficult to directly evaluate $\delta G^{WL} (T)$ within 
this technique.

One of the problems in those cases is that the description in
terms of quasiclassical electron trajectories may become
insufficient, and electron scattering on disorder should be
treated on more general footing. Another (though purely technical)
point is averaging over disorder. In our approach \cite{GZ}
disorder averaging is (can be) postponed until the last stage of
the calculation which is convenient in certain physical
situations. In other cases -- like ones studied below -- it might
be, in contrast, more appropriate to perform disorder averaging
already in the beginning of the whole analysis. In addition, it is
desirable to deal with the model which would embrace various types
of conductors with well defined properties both in the long and short
wavelength limits.

In this paper we make a first step towards this unified theory.
Namely, we will describe a disordered conductor by means of an
array of (metallic) quantum dots connected via junctions
(scatterers) with an arbitrary distribution of transmissions of
their conducting channels. This model will allow to easily
crossover between the limits of granular metals and those with
point-like impurities and to treat spatially restricted and
spatially extended conductors within the same theoretical
framework. Electron scattering on each such scatterer will be
treated within the most general scattering matrix formalism
\cite{MB,B} adopted to include electron-electron interaction
effects \cite{Naz,GZ00,BN,GZ041,GZ042}. Averaging over disorder
will be performed within the non-linear $\sigma-$model technique
in Keldysh formulation as first proposed by Horbach and Sch\"on
\cite{HS} for non-interacting electrons. This method
has certain advantages over the imaginary time approach
\cite{efetov1} since it allows to treat both equilibrium and
non-equilibrium problems and also enables one to include Coulomb
interaction between electrons in a straightforward manner
\cite{KA}. In this paper we will analyze WL corrections to
conductance merely for non-interacting electrons and will include
interaction effects only phenomenologically by introducing an
effective electron dephasing time $\tau_\varphi$ as a parameter of
our theory. Systematic analysis of the effect of
electron-electron interactions on weak localization within this 
formalism will be developed elsewhere.

The structure of the paper is as follows. In Sec. II we define the basic model
of a 1d array of quantum dots and outline the key features of our
formalism. In Sec. III we will introduce convenient parameterization of the
non-linear $\sigma$-model which will then be used in Sec. IV to derive WL
correction to the system conductance for the model in question. This WL
correction will be evaluated for various structures in Sec. V. Sec. VI
contains direct generalization of our analysis and results to the case
of 2d arrays of quantum dots and is followed by a brief summary in Sec. VII.
Some technical details of our calculation are presented in Appendix.

\section{The model and formalism}

\begin{figure}
\centerline{\includegraphics[width=9cm]{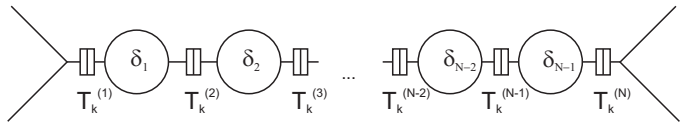}} \caption{1d
array of $N-1$ quantum dots coupled by $N$ barriers. Each quantum
dot is characterized by mean level spacing $\delta_n$. Each
barrier is characterized by a set of transmissions of its
conducting channels $T_k^{(n)}$.} \label{array1}
\end{figure}

Let us consider a 1d array of connected in series chaotic quantum dots
(Fig. \ref{array1}). Each quantum dot is characterized by its own
mean level spacing $\delta_n$. Adjacent quantum dots are connected
via barriers which can scatter electrons. Each such scatterer is
described by a set of transmissions of its conducting channels
$T_k^{(n)}$ (here $k$ labels the channels and $n$ labels the
scatterers). We will ignore spin-orbit scattering and,
for the sake of definiteness and simplicity, we will
first focus our attention on 1d arrays only. Generalization of our
analysis to other situations can be performed in a straightforward
manner, as it will be demonstrated in Sec. VI of the paper.

An effective action $S[\check Q]$ of an array depicted in
Fig. \ref{array1} depends on
the fluctuating $4\times 4$ matrix fields \cite{SLF} $\check Q_n(t_1,t_2)$ defined
for each of the dots ($n=1,...,N-1$). Each of these fields
is a function of two times $t_1$ and $t_2$ and obeys
the normalization condition
\begin{equation}
\check Q_n^2=1.
\label{norm}
\end{equation}
The action of an array can be represented as a sum of two terms
\begin{eqnarray}
iS[\check Q]=iS_{d}[\check Q]+iS_{t}[\check Q].
\label{action}
\end{eqnarray}
The first term, $iS_d[\check Q]$, describes the contribution of
bulk parts of the dots. This term reads
\begin{eqnarray}
iS_d[\check Q]=\sum_{n=1}^{N-1}\frac{\pi}{\delta_n}\,{\rm Tr}\,
\left[\frac{\partial}{\partial t}\check Q_n-\alpha_n H^2\big([\check A,\check Q_n]\big)^2\right].
\label{Sd}
\end{eqnarray}
Here $H$ is an external magnetic filed,
$\alpha_n= b_n (e^2/ \hbar^2c^2)v_Fd_n^2\min\{l_e,d_n\}$, 
$b_n$ is a geometry dependent numerical
prefactor \cite{efetov,B}, $d_n$ is the size
of $n-$th dot, $l_e$ is the elastic mean free path in
the dot, and $\check A$ is $4\times 4$ matrix:
\begin{eqnarray}
\check A=\left(
\begin{array}{cccc}
1 & 0 & 0 & 0 \\
0 & -1 & 0 & 0 \\
0 & 0 & 1 & 0 \\
0 & 0 & 0 & -1
\end{array}
\right).
\end{eqnarray}
The second term  in Eq. (\ref{action}), $iS_t[\check Q]$, describes electron
transfer between quantum dots. It has the form \cite{Nazarov}
\begin{eqnarray}
iS_t[\check Q]=\frac{1}{2}\sum_{n=1}^N\sum_{k}\,{\rm Tr}\,\ln
\left[1+\frac{T^{(n)}_k}{4}\big(\{\check Q_{n-1},\check Q_n\}-2\big)\right].
\label{Sj}
\end{eqnarray}
A similar expression was also considered within the imaginary time
technique \cite{IWZ,efetov}.

Note that here the magnetic field $H$ is included only in the term
(\ref{Sd}) describing the quantum dots while it is ignored in the
term (\ref{Sj}). Usually this approximation remains applicable at
not too low magnetic fields. We will return to this point in Sec.
VI.

An equilibrium saddle point configuration $\check\Lambda(t_1-t_2)$ 
of the matrix field
$\check Q(t_1,t_2)$ depends only on the time difference and has the form
\begin{eqnarray}
\check \Lambda(t)=\int\frac{dE}{2\pi}\, e^{-iEt}\left(
\begin{array}{cccc}
-1 & 0 & 0 & 0 \\
0 & 1 & 0 & 0 \\
g^K(E) & 0 & 1 & 0 \\
0 & -g^K(E) & 0 & -1
\end{array}
\right),
\end{eqnarray}
where $g^K(E)=2[1-2f_F(E)]=2\tanh(E/2T).$
This choice of the saddle point corresponds to the following
structure of the $4\times 4$ matrix Green function $\check G$:
\begin{eqnarray}
\check G=\left(
\begin{array}{cccc}
G^A & 0 & 0 & 0 \\
0 & {\cal T}G^{A*}{\cal T} & 0 & \\
-G^K & 0 & G^R & 0 \\
0 &  {\cal T}G^{K*}{\cal T} & 0 & {\cal T}G^{R*}{\cal T}
\end{array}
\right).
\label{G}
\end{eqnarray}
Here we defined the time inversion operator ${\cal T}$:
\begin{eqnarray}
{\cal T} f(t)=f(t_f-t),
\label{T}
\end{eqnarray}
 where $t_f$ will be specified later. Note that the function
$\check G$ in Eq. (\ref{G}), defined for a given disorder configuration,
should be contrasted from the Green function
\begin{eqnarray}
\check G_Q=\left[i\frac{\partial}{\partial t}+\frac{\nabla^2}{2m}
+\frac{i}{2\tau_e}\check Q\right]^{-1}
\label{GQ}
\end{eqnarray}
defined for a given realization of the matrix field $\check Q$.
In Eq. (\ref{GQ}) we also introduced the electron elastic mean free time $\tau_e$.

\section{Gaussian approximation}

In order to evaluate the WL correction to conductance we will
account for quadratic (Gaussian) fluctuations of the matrix field
$\check Q_n$. This approximation is always sufficient provided 
the conductance of the whole sample exceeds  $e^2/h$, in certain situations
somewhat softer applicability conditions can be formulated. Expanding in powers of such fluctuations we introduce
the following parameterization
\begin{eqnarray}
&& \check Q_n=e^{i\check W_n}\check\Lambda e^{-i\check W_n}
\nonumber\\ &&
=\check\Lambda +i[\check W_n,\check \Lambda]+\check W_n\check\Lambda\check W_n
-\frac{1}{2}\{\check W^2_n,\check\Lambda\}+{\cal O}(W^3).\;\;\;\;\;\;
\label{Q1}
\end{eqnarray}
It follows from the normalization condition (\ref{norm}) that only
$8$ out of $16$ matrix elements of $\check W$ are independent
parameters. This observation provides certain freedom to choose an
explicit form of of this matrix. A convenient parameterization to
be used below is
\begin{eqnarray}
\check W_n=\left(
\begin{array}{cccc}
0 & u_{1n} & b_{1n} & 0 \\
u_{2n} & 0 & 0 & b_{2n} \\
a_{1n}+b_{1n} & 0 & 0 & v_{1n} \\
0 & a_{2n}+b_{2n} & v_{2n} & 0
\end{array}
\right).
\end{eqnarray}
With this choice the quadratic part of the action takes the form
\begin{eqnarray}
iS^{(2)}=iS^{(2)}_{ab}[a,b]+iS^{(2)}_{uv}[u,v],
\end{eqnarray}
where $iS^{(2)}_{ab}[a,b]$ does not depend on $H$
and describes diffuson modes, while $iS^{(2)}_{uv}[u,v]$ is sensitive to the
magnetic field and is responsible for the Cooperons.
The diffuson part of the action $iS^{(2)}_{ab}[a,b]$ was already analyzed in
Ref. \onlinecite{GZ042} and will be omitted here. Below we will focus our
attention on the Cooperon contribution which reads
\begin{eqnarray}
iS^{(2)}_{uv}[u,v]&=& \sum_{n=1}^{N-1}\frac{2\pi}{\delta_n}\,{\rm Tr}\,
\bigg[ \frac{\partial}{\partial t}[u_{1n},u_{2n}]-16\alpha_nH^2\, u_1u_2\bigg]
\nonumber\\ &&
+\,\sum_{n=1}^{N-1}\frac{2\pi}{\delta_n}\,{\rm Tr}\,
\bigg[ \frac{\partial}{\partial t}[v_{2n},v_{1n}]-16\alpha_nH^2\, v_1v_2\bigg]
\nonumber\\ &&
-\,\sum_{n=1}^{N}\frac{g_n}{2}\,{\rm Tr}\,\bigg[
(u_{1n}-u_{1,n-1})(u_{2n}-u_{2,n-1})
\nonumber\\ &&
+\,(v_{1n}-v_{1,n-1})(v_{2n}-v_{2,n-1})
\bigg],
\label{quadr}
\end{eqnarray}
where $g_n=2\sum_k T_k^{(n)}=2\pi\hbar/e^2R_n$ is the dimensionless conductance
of $n-$th barrier. With the aid of the action (\ref{quadr}) we can derive
the pair correlators of the fields $u_{1,2}$ and $v_{1,2}$:
\begin{eqnarray}
\langle u_{1n}(t_1,t_2)u_{2m}(t',t'')\rangle=
\langle v_{1n}(t',t'')v_{2m}(t_1,t_2)\rangle
\nonumber\\
=\frac{\delta_m}{2\pi}\delta(t_1-t_2+t'-t'')
C_{nm}(t''-t_1),
\label{avuv}
\end{eqnarray}
where we defined a discrete version of the Cooperon $C_{nm}(t)$
obeying the equation
\begin{eqnarray}
\left(\frac{\partial}{\partial t}+\frac{1}{\tau_{Hn}}+\frac{1}{\tau_{\varphi n}}\right)C_{nm}
+\frac{\delta_n}{4\pi}\big[(g_n+g_{n+1})C_{nm}
\nonumber\\
-\,g_nC_{n-1,m}-g_{n+1}C_{n+1,m}\big]
=\delta_{nm}\delta(t).
\label{diff}
\end{eqnarray}
This equation should be supplemented by the boundary condition
$C_{nm}(t)=0$ which applies whenever one of the indices
$n$ or $m$ belongs to the lead electrode. Here $\tau_{Hn}=1/16\alpha_nH^2$ is
the electron dephasing time due to the magnetic field. In Eq. (\ref{diff})
we also introduced an additional electron decoherence time in $n-$th quantum dot $\tau_{\varphi n}$
which can remain finite in the presence of interactions. In this paper we
are not aiming to further specify the interaction mechanisms and only
account for them phenomenologically by keeping the parameter
$\tau_{\varphi n}$ in the equation for the Cooperon.

\section{WL corrections}

Let us now derive an expression for WL correction to the
conductance in terms of the fluctuating fields $u$ and $v$. In
what follows we will explicitly account for the discrete nature of
our model and specify the WL correction for a single barrier
in-between two adjacent quantum dots in the array.

We start, however, from the bulk limit, in which case
the Kubo formula for the conductivity tensor $\sigma_{\alpha\beta}$
reads
\begin{eqnarray}
&& \sigma_{\alpha\beta}(\bm{r},\bm{r}')=-i\int_{-\infty}^t dt'\; (t-t')
\nonumber\\ &&\times\,
\langle j_\beta(t',\bm{r}')j_{\alpha}(t,\bm{r})-j_{\alpha}(t,\bm{r})j_\beta(t',\bm{r}') \rangle.
\label{Kubo}
\end{eqnarray}
Following the standard procedure \cite{Berg,AA,CS}, approximating
the Fermi function as $-\partial f_F(E)/\partial E\approx\delta(E)$ (which
effectively implies taking the low temperature limit)
and using a phenomenological description of interactions as mediated by 
external (classical) fluctuating fields, from Eq. (\ref{Kubo}) one can 
derive the WL correction in the form:
\begin{eqnarray}
&&\delta \sigma_{\alpha\beta}^{WL}(\bm{r},\bm{r}')=-\frac{e^2}{4\pi m^2}\int_{-\infty}^t dt'\int dt''
\nonumber\\ &&\times\,
(\nabla_{\bm{r}_1}^\alpha - \nabla_{\bm{r}_2}^\alpha)_{\bm{r}_1=\bm{r}_2=\bm{r}}
(\nabla_{\bm{r}'_1}^\beta - \nabla_{\bm{r}'_2}^\beta)_{\bm{r}'_1=\bm{r}'_2=\bm{r}'}
\nonumber\\ && \times\,
\left\langle
G^R(t,\bm{r}_1; t'',\bm{r}'_2)G^A(t',\bm{r}'_1; t,\bm{r}_2)
\right\rangle_{\rm dis,\; max\; cross},\hspace{0.5cm}
\label{sigma1}
\end{eqnarray}
which implies summation over all maximally crossed
diagrams\cite{Berg,AA,CS}, as indicated in the subscript.
At the same time, averaging over fluctuations of $\check Q$
within Gaussian approximation is equivalent to summing
over all ladder diagrams. Since we are not going to go beyond
the above approximation, we need to convert maximally crossed diagrams
in Eq. (\ref{sigma1}) into the ladder ones. Technically this conversion
can be accomplished by an effective time reversal procedure 
for the advanced Green function which can be illustrated as follows. 

Consider, e. g., the second order correction to $G^A$ in the disorder potential $U_{\rm dis}(\bm{x})$
\begin{eqnarray}
&&\delta^{(2)} G^A(t',\bm{r}'_1; t,\bm{r}_2)= -i\int_{t'}^t d\tau_2\int_{t'}^{\tau_2}d\tau_1
\int d^3\bm{x_2}d^3\bm{x_1}
\nonumber\\ &&\times\,
 G^A(t',\bm{r}'_1; \tau_1,\bm{x}_1)U_{\rm dis}(\bm{x}_1)G^A(\tau_1,\bm{x}_1; \tau_2,\bm{x}_2)
\nonumber\\ &&\times\,
U_{\rm dis}(\bm{x}_2)G^A(\tau_2,\bm{x}_2; t,\bm{r}_2).
\end{eqnarray}
Making use of the property $G^A(X_1,X_2)=G^{R*}(X_2,X_1)$, we get
\begin{eqnarray}
&&\delta^{(2)} G^A(t',\bm{r}'_1; t,\bm{r}_2)= -i\int_{t'}^t d\tau_2\int_{t'}^{\tau_2}d\tau_1
\int d^3\bm{x_2}d^3\bm{x_1}
\nonumber\\ &&\times\,
 G^{R*}(t,\bm{r}_2; \tau_2,\bm{x}_2)U_{\rm dis}(\bm{x}_2)G^{R*}(\tau_2,\bm{x}_2; \tau_1,\bm{x}_1)
\nonumber\\ &&\times\,
U_{\rm dis}(\bm{x}_1)G^{R*}(\tau_1,\bm{x}_1; t',\bm{r}'_1).
\end{eqnarray}
Setting $t_f=t+t'$,
we rewrite this expression as follows
\begin{eqnarray}
&&\delta^{(2)} G^A(t',\bm{r}'_1; t,\bm{r}_2)= -i\int_{t_f-t}^{t_f-t'} d\tau_2\int_{t_f-t}^{\tau_2}d\tau_1
\nonumber\\ &&\times\,
\int d^3\bm{x_2}d^3\bm{x_1}\;G^{R*}(t_f-t',\bm{r}_2; \tau_2,\bm{x}_2)
\nonumber\\ &&\times\,
 U_{\rm dis}(\bm{x}_2)G^{R*}(\tau_2,\bm{x}_2; \tau_1,\bm{x}_1)
\nonumber\\ &&\times\,
U_{\rm dis}(\bm{x}_1)G^{R*}(\tau_1,\bm{x}_1; t_f-t,\bm{r}'_1).
\label{20}
\end{eqnarray}
Close inspection of the right hand side of Eq. (\ref{20}) allows 
to establish the following relation
\begin{eqnarray}
\delta^{(2)} G^A(t',\bm{r}'_1; t,\bm{r}_2)={\cal T}\delta^{(2)}G^{R*}(t',\bm{r}_2; t,\bm{r}'_1){\cal T},
\end{eqnarray}
which turns out to hold in all orders of the perturbation theory in $U_{\rm
  dis}$. As before, the time inversion operator ${\cal T}$ is defined in Eq. (\ref{T})
with $t_f=t+t'$.

As a result, the expression for $\delta \sigma_{\alpha\beta}^{WL}$ takes the form:
\begin{eqnarray}
&& \delta \sigma_{\alpha\beta}^{WL}(\bm{r},\bm{r}')=-\frac{e^2}{4\pi m^2}\int_{-\infty}^t dt'\int dt''
\nonumber\\ &&\times\,
(\nabla_{\bm{r}_1}^\alpha - \nabla_{\bm{r}_2}^\alpha)_{\bm{r}_1=\bm{r}_2=\bm{r}}
(\nabla_{\bm{r}'_1}^\beta - \nabla_{\bm{r}'_2}^\beta)_{\bm{r}'_1=\bm{r}'_2=\bm{r}'}
\nonumber\\ && \times\,
\left\langle
G^R(t,\bm{r}_1; t'',\bm{r}'_2){\cal T}G^{R*}(t',\bm{r}_2; t,\bm{r}'_1){\cal T}
\right\rangle_{\rm dis,\; ladder}\;\;\;\;\;
\label{sigma2}
\end{eqnarray}
Rewriting Eq. (\ref{sigma2})
in terms of the matrix elements of the Green function (\ref{G}), we obtain
\begin{eqnarray}
&& \delta \sigma_{\alpha\beta}^{WL}(\bm{r},\bm{r}')=-\frac{e^2}{4\pi m^2}\int_{-\infty}^t dt'\int dt''
\nonumber\\ &&\times\,
(\nabla_{\bm{r}_1}^\alpha - \nabla_{\bm{r}_2}^\alpha)_{\bm{r}_1=\bm{r}_2=\bm{r}}
(\nabla_{\bm{r}'_1}^\beta - \nabla_{\bm{r}'_2}^\beta)_{\bm{r}'_1=\bm{r}'_2=\bm{r}'}
\nonumber\\ && \times\,
\left\langle
G_{33}(t,\bm{r}_1; t'',\bm{r}'_2)G_{44}(t',\bm{r}_2; t,\bm{r}'_1)
\right\rangle_{\rm dis,\; ladder}
\label{sigma3}
\end{eqnarray}

Our next step amounts to expressing WL correction via the Green
function $\check G_Q$ (\ref{GQ}). For that purpose
we will use the following rule of averaging
\begin{eqnarray}
&& \left\langle
G_{33}(t,\bm{r}_1; t'',\bm{r}'_2)G_{44}(t',\bm{r}_2; t,\bm{r}'_1)
\right\rangle_{\rm dis}
\nonumber\\ &&
=\left\langle
G_{33;Q}(t,\bm{r}_1; t'',\bm{r}'_2)G_{44;Q}(t',\bm{r}_2; t,\bm{r}'_1)
\right\rangle_{Q}
\nonumber\\ &&
-\, \left\langle
G_{34;Q}(t,\bm{r}_1;t,\bm{r}'_1 )G_{43;Q}(t',\bm{r}_2;  t'',\bm{r}'_2)
\right\rangle_{Q}.
\label{av}
\end{eqnarray}
One can check that within our Gaussian approximation in $u$ and
$v$ the first term in the right hand side of Eq. (\ref{av}) does
not give any contribution. Hence, we find
\begin{eqnarray}
&& \delta \sigma_{\alpha\beta}^{WL}(\bm{r},\bm{r}')=\frac{e^2}{4\pi m^2}\int_{-\infty}^t dt'\int dt''
\nonumber\\ &&\times\,
(\nabla_{\bm{r}_1}^\alpha - \nabla_{\bm{r}_2}^\alpha)_{\bm{r}_1=\bm{r}_2=\bm{r}}
(\nabla_{\bm{r}'_1}^\beta - \nabla_{\bm{r}'_2}^\beta)_{\bm{r}'_1=\bm{r}'_2=\bm{r}'}
\nonumber\\ && \times\,
\left\langle
G_{34; Q}(t,\bm{r}_1; t,\bm{r}'_1)G_{43; Q}(t',\bm{r}_2; t'',\bm{r}'_2)
\right\rangle_{Q}.
\label{sigma4}
\end{eqnarray}

Let us now turn to our model of Fig. 1 in which case the voltage
drops occur only across barriers. In this case Eq. (\ref{sigma4}),
which only applies to bulk metals, should be generalized
accordingly. Consider the conductance of an individual barrier
determined by the following Kubo formula
\begin{eqnarray}
G&=&-i\int_{-\infty}^t dt' (t-t')
\langle I(t',x')I(t,x)
\nonumber\\ &&
-\,I(t,x)I(t',x') \rangle.
\label{KuboG}
\end{eqnarray}
Here $I(t,x)$ is the operator of the total current flowing in the lead (or dot)
and $x$ is a longitudinal coordinate chosen to be in a close vicinity of the barrier. Due to the current conservation the
conductance $G$ should not explicitly depend on $x$ and $x'$.
Comparing
Eqs. (\ref{KuboG}) and (\ref{Kubo}), and making use of Eq. (\ref{sigma4})
and the relation $I(t,x)=\int d^2{\bm z}\, j_x(t,x,\bm{z}),$
where $j_x$ is the current density in the $x-$direction and $\bm{z}$ is the vector in the transversal
direction, we conclude that WL correction to the conductance of a
barrier between the left and right dots should read
\begin{eqnarray}
&& \delta G^{WL}_{LR}=\frac{e^2}{4\pi m^2}\int_{-\infty}^t dt'\int dt''\int d^2\bm{z}d^2\bm{z}'
\nonumber\\ &&\times\,
(\nabla_{x_1} - \nabla_{x_2})_{x_1=x_2=x}
(\nabla_{x'_1} - \nabla_{x'_2})_{x'_1=x'_2=x'}
\nonumber\\ && \times\,
\left\langle
G_{34; Q}(t,x_1,\bm{z}; t,x'_1,\bm{z}')G_{43; Q}(t',x_2,\bm{z}; t'',x'_2,\bm{z'})
\right\rangle_{Q}.
\nonumber\\
\label{GWL1}
\end{eqnarray}

In what follows we will assume that both coordinates
$x$ and $x'$ are on the left side from and very close to the
corresponding barrier. Let us express the Green function
in the vicinity of the barrier in the form
\begin{eqnarray}
&& \check G_Q(t,x,\bm{z};t',x',\bm{z}')=\sum_{nm}\big\{
e^{ip_nx_1-ip_mx'}\check {\cal G}^{++}_{mn}(t,t',x,x')
\nonumber\\ &&
+\,e^{-ip_nx+ip_mx'}\check {\cal G}^{--}_{mn}(t,t',x,x')
\nonumber\\ &&
+\,e^{ip_nx+ip_mx'}\check {\cal G}^{+-}_{mn}(t,t',x,x')
\nonumber\\ &&
+\,e^{-ip_nx-ip_mx'}\check {\cal G}^{-+}_{mn}(t,t',x,x')\big\}
\Phi_n(\bm{z})\Phi_m^*(\bm{z}'),
\end{eqnarray}
where $\Phi_n(\bm{z})$ are the transverse quantization modes
which define  conducting channels, $p_n$ is projection of the Fermi
momentum perpendicular to the surface of the barrier, and
the semiclassical Green function ${\cal G}_{mn}^{\alpha\beta}$ slowly varies in space.
 Eq. (\ref{GWL1}) then becomes
\begin{eqnarray}
&& \delta G^{WL}_{LR}=\frac{e^2}{4\pi m^2}\int_{-\infty}^t dt'\int dt''
\nonumber\\ &&\times\,
\sum_{mnkl}\sum_{\alpha\beta\gamma\delta=\pm 1}
(\alpha p_n-\gamma p_k)(\beta p_m-\delta p_l)
\nonumber\\ && \times\,
\left\langle
{\cal G}_{mn;34}^{\alpha\beta}(t,t,x,x'){\cal G}_{kl;43}^{\gamma\delta}(t',t'',x,x')
\right\rangle_{Q}
\nonumber\\ &&\times\,
\left. e^{i\alpha p_nx_1-i\beta p_mx'_1+i\gamma p_kx_2- i\delta p_lx'_2}
\right|_{x_1=x_2=x;x'_1=x'_2=x'}.\hspace{0.75cm}
\label{GWL2}
\end{eqnarray}

Next we  require $\delta G^{WL}_{LR}$ to be independent on $x$ and $x'$, i.e.
in Eq. (\ref{GWL2}) we omit those terms, which contain
quickly oscillating functions of these coordinates.
This requirement implies that $\alpha p_n+\gamma p_k=0$
and $\beta p_m+\delta p_l=0$. These constraints in turn yield $\gamma=-\alpha$, $\delta=-\beta$,
$k=n$ and $l=m$. Thus, we get
\begin{eqnarray}
\delta G^{WL}_{LR}=\frac{e^2}{\pi m^2}\sum_{mn}\sum_{\alpha\beta=\pm 1}\int_{-\infty}^t dt'\int dt''
\alpha\beta p_n p_m
\nonumber\\  \times\,
\left\langle
{\cal G}_{mn;34}^{\alpha\beta}(t,t,x,x'){\cal G}_{nm;43}^{-\alpha,-\beta}(t',t'',x,x')
\right\rangle_{Q}.
\label{GWL3}
\end{eqnarray}

Let us choose the basis in which transmission and reflection
matrices $\hat t$ and $\hat r$ are diagonal. In this basis the
semiclassical Green function is diagonal as well, ${\cal
G}_{mn}\propto {\cal G}_{nn}\delta_{nm}$, and Eq. (\ref{GWL3})
takes the  form
\begin{eqnarray}
\delta G^{WL}_{LR}&=&\frac{e^2}{\pi}\sum_{n}\frac{p^2_n}{m^2}\int_{-\infty}^t dt'\int dt''
\nonumber\\ && \times\,
\big\langle
{\cal G}_{L,nn;34}^{++}(t,t){\cal G}_{L,nn;43}^{--}(t',t'')
\nonumber\\ &&
+\,{\cal G}_{L,nn;34}^{--}(t,t){\cal G}_{L,nn;43}^{++}(t',t'')
\nonumber\\ &&
-\,{\cal G}_{L,nn;34}^{+-}(t,t){\cal G}_{L,nn;43}^{-+}(t',t'')
\nonumber\\ &&
-\,{\cal G}_{L,nn;34}^{-+}(t,t){\cal G}_{L,nn;43}^{+-}(t',t'')
\big\rangle_{Q}.
\label{GWL4}
\end{eqnarray}
What remains  is to express WL correction  in terms of the field
$\check Q$ only. This goal is achieved by establishing an explicit
relation between the Green function $\check {\cal G}$ and the
field $\check Q$. A derivation of this relation is presented  in
Appendix A. Here we only display the final result expressed via
the fluctuating fields $v_1$ and $v_2$. We obtain
\begin{eqnarray}
&& \delta G^{WL}_{LR}=-\frac{e^2}{\pi }\sum_{n}\int_{-\infty}^t dt'\int dt''
\nonumber\\ &&  \times\,
\big\langle
T_n \big[ v_{1L}(t,t)v_{2R}(t',t'')+v_{1R}(t,t)v_{2L}(t',t'') \big]
\nonumber\\ &&
+\, T_n^2 [v_{1L}(t,t)-v_{1R}(t,t)][v_{2L}(t',t'')-v_{2R}(t',t'')]
\big\rangle.\hspace{0.65cm}
\label{GWL5}
\end{eqnarray}
Note that the contribution linear in $T_n$, which
contains the product of the fluctuating
fields on two different sides of the barrier, vanishes identically
provided fluctuations on one side tend to zero, e.g. if the barrier is
directly attached to a large metallic lead. In contrast, the contribution
$\propto T_n^2$ in Eq. (\ref{GWL5}) survives even in this case.

Finally, applying the contraction rule (\ref{avuv}) we get
\begin{eqnarray}
 \delta G^{WL}_{LR}&=&-\frac{e^2 g}{4\pi^2 }\int_{0}^\infty dt
\big\{ \beta\big[\delta_R C_{LR}(t)+\delta_L C_{RL}(t)\big]
\nonumber\\ &&
+\,(1-\beta)\big[\delta_R C_{RR}(t)+\delta_LC_{LL}(t)\big]\big\}.
\label{GWL}
\end{eqnarray}
Here $\delta_{L,R}$ is the mean level spacing in the left/right quantum dot,
$g=2\sum_k T_k$ is the dimensionless conductance of the barrier and
 $\beta=\sum_kT_k(1-T_k)/\sum_k T_k$ is the corresponding Fano factor.

Likewise, the WL correction to the $n-$th barrier conductance in
1d array of $N-1$ quantum dots with mean level spacings $\delta_n$
connected by $N$ barriers with dimensionless conductances $g_n$
and Fano factors $\beta_n$ reads
\begin{eqnarray}
\delta G^{WL}_n&=&-\frac{e^2 g_n}{4\pi^2 }\int_{0}^\infty dt
\big\{ \beta_n\big[\delta_n C_{n-1,n}(t)
\nonumber\\ &&
+\,\delta_{n-1} C_{n,n-1}(t)\big]+(1-\beta_n)\big[\delta_n C_{nn}(t)
\nonumber\\ &&
+\,\delta_{n-1}C_{n-1,n-1}(t)\big]\big\}.
\label{GWLn}
\end{eqnarray}

So far we discussed the local properties, namely WL corrections to
the conductivity tensor, $\delta \sigma^{WL}_{\alpha,\beta}(\bm{r},\bm{r}'),$ and
to the conductance of a single barrier, $\delta G^{WL}_{LR}$. Our main goal
is, however, to evaluate the WL correction to the conductance of the whole
system. For bulk
metals one finds that at large scales the WL correction (\ref{sigma1}) is local,
$\delta \sigma^{WL}_{\alpha,\beta}(\bm{r},\bm{r}')\propto \delta(\bm{r}-\bm{r}').$
In general though, there can exist other, non-local, contributions to the
conductivity tensor \cite{Kane}. Without going into details here, we only
point out that, even if these non-local terms are present, 
one can still apply the standard Ohm's law arguments in order to obtain 
the conductance of the whole sample. Specifically, in the case of 1d arrays one finds
(cf. \cite{Argaman1}): 
\begin{eqnarray}
\delta G^{WL}&=&\frac{1}{\sum_{n=1}^N (G_n+\delta G^{WL}_n)^{-1}} -
\frac{1}{\sum_{n=1}^N G_n^{-1}}
\nonumber\\
&=& \frac{\sum_{n=1}^N \delta G^{WL}_n/g_n^2}{\left(\sum_{n=1}^N 1/g_n\right)^2} +
{\rm higher\; order\; terms}.\hspace{0.5cm}
\label{Garray}
\end{eqnarray}
Eqs. (\ref{GWL}), (\ref{GWLn}) and (\ref{Garray})  will be used to evaluate WL
corrections for different configurations of quantum dots considered below.

\section{Examples}

\subsection{Single quantum dot}

\begin{figure}
\centerline{\includegraphics[width=5cm]{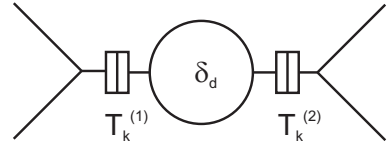}} \caption{Single
quantum dot connected to the leads via two barriers.}
\end{figure}
We start from the simplest case of a single quantum dot depicted in
Fig. 2. In this case the solution of Eq. (\ref{diff})
reads
\begin{eqnarray}
C_{11}(t)=\exp\left[-\frac{t}{\tau_D}-\frac{t}{\tau_H}-\frac{t}{\tau_\varphi}\right],
\end{eqnarray}
where $\tau_D=4\pi / (g_1+g_2)\delta_d$ is the dwell time,
and $\delta_d$ is the
mean level spacing in the quantum dot. All other components of
the Cooperon are equal to zero. From Eq. (\ref{GWL}) we get
\begin{eqnarray}
\delta G^{WL}_1&=&-\frac{e^2 g_1(1-\beta_1)\delta_d}{4\pi^2}\frac{1}{1/\tau_D+1/\tau_{H}+1/\tau_\varphi},
\nonumber\\
\delta G^{WL}_2&=&-\frac{e^2 g_2(1-\beta_2)\delta_d}{4\pi^2}\frac{1}{1/\tau_D+1/\tau_{H}+1/\tau_\varphi}.
\end{eqnarray}
According to Eq. (\ref{Garray}) the total WL correction
becomes
\begin{eqnarray}
\delta G^{WL}=-\frac{e^2\delta}{4\pi^2}
\frac{g_1g_2^2(1-\beta_1)+g_1^2g_2(1-\beta_2)}{(g_1+g_2)^2
\left({1}/{\tau_D}+{1}/{\tau_\varphi}+{1}/{\tau_H}\right)}.
\label{magres1qd}
\end{eqnarray}
Since $1/\tau_H\propto H^2$, the magnetoconductance has the
Lorentzian shape\cite{B}. In the limit $H=0$ and in the absence of
interactions ($\tau_\varphi\to\infty$) Eq. (\ref{magres1qd})
reduces to \cite{BB}
\begin{eqnarray}
\delta G^{WL}=-\frac{e^2}{\pi}
\frac{g_1g_2^2(1-\beta_1)+g_1^2g_2(1-\beta_2)}{(g_1+g_2)^3}.
\label{1qd}
\end{eqnarray}

\subsection{Two quantum dots}

\begin{figure}
\centerline{\includegraphics[width=3cm]{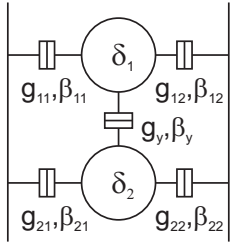}}
\caption{Most general system with two quantum dots}
\label{twodotgen}
\end{figure}

Next we consider the most general setup composed of two quantum
dots with the corresponding conductances and Fano factors defined
as in Fig. \ref{twodotgen}. The Cooperon is represented as a
$2\times 2$ matrix which zero frequency component satisfies the
following equation
\begin{eqnarray}
\left(\begin{array}{c}
g_{11}+g_{12}+g_y+\gamma_1 \hspace{1cm} -g_y \\
-g_y \hspace{1cm} g_{21}+g_{22}+g_y+\gamma_2
\end{array}\right)
\left(
\begin{array}{cc}
C_{11} & C_{12} \\
C_{21} & C_{22}
\end{array}\right)
\nonumber\\
=\,\left(
\begin{array}{cc}
4\pi/\delta_1 & 0 \\
0 & 4\pi/\delta_2
\end{array}
\right),
\end{eqnarray}
where
\begin{eqnarray}
\gamma_{1,2}=\frac{4\pi}{\delta_{1,2}}
\left(\frac{1}{\tau_{H1,2}}+\frac{1}{\tau_{\varphi 1,2}}\right).
\label{gamma}
\end{eqnarray}
Defining $\Delta=(g_{11}+g_{12}+g_y+\gamma_1)(g_{21}+g_{22}+g_y+\gamma_2)-g_y^2$, we get
\begin{eqnarray}
\left(
\begin{array}{cc}
C_{11} & C_{12} \\
C_{21} & C_{22}
\end{array}\right)=
\frac{4\pi}{\Delta}\left(
\begin{array}{c}
(g_{21}+g_{22}+g_y+\gamma_2)/\delta_1 \hspace{0.2cm} g_y/\delta_2 \\
g_y/\delta_1 \hspace{0.2cm} (g_{11}+g_{12}+g_y+\gamma_1)/\delta_2
\end{array}
\right).
\nonumber
\end{eqnarray}
With the aid of Eq. (\ref{GWL}) we derive WL corrections for all
five barriers in our setup:
\begin{eqnarray}
 \delta G_{11}^{WL}&=&-\frac{e^2g_{11}\delta_1(1-\beta_{11})}{4\pi^2}C_{11}
\nonumber\\ &&
=\,-\frac{e^2}{\pi}\frac{g_{11}(g_{21}+g_{22}+g_y+\gamma_2)(1-\beta_{11})}{\Delta},
\nonumber\\
\delta
G_{12}^{WL}&=&-\frac{e^2g_{12}\delta_1(1-\beta_{12})}{4\pi^2}C_{11}
\nonumber\\ &&
=\,-\frac{e^2}{\pi}\frac{g_{12}(g_{21}+g_{22}+g_y+\gamma_2)(1-\beta_{12})}{\Delta},
\nonumber\\
\delta
G_{21}^{WL}&=&-\frac{e^2g_{21}\delta_1(1-\beta_{21})}{4\pi^2}C_{22}
\nonumber\\ &&
=\,-\frac{e^2}{\pi}\frac{g_{21}(g_{11}+g_{12}+g_y+\gamma_1)(1-\beta_{21})}{\Delta},
\nonumber\\
 \delta G_{22}^{WL}&=&-\frac{e^2g_{22}\delta_1(1-\beta_{22})}{4\pi^2}C_{22}
\nonumber\\ &&
=\,-\frac{e^2}{\pi}\frac{g_{22}(g_{11}+g_{12}+g_y+\gamma_1)(1-\beta_{22})}{\Delta},
\nonumber\\
 \delta G_{y}^{WL}&=&-\frac{e^2g_{y}}{4\pi^2}\big[ \beta_y(\delta_1 C_{21}+\delta_2 C_{12})
\nonumber\\ &&
+\,(1-\beta_y)(\delta_1C_{11}+\delta_2C_{22}) \big]
\nonumber\\
&=&-\frac{e^2g_y}{\pi\Delta}\big[
2g_y\beta_y+(1-\beta_y)(g_{11}+g_{12}+g_{21}
\nonumber\\ &&
+\,g_{22}+2g_y +\gamma_1+\gamma_2)
\big].
\label{Gij}
\end{eqnarray}

\begin{figure}
\centerline{\includegraphics[width=8cm]{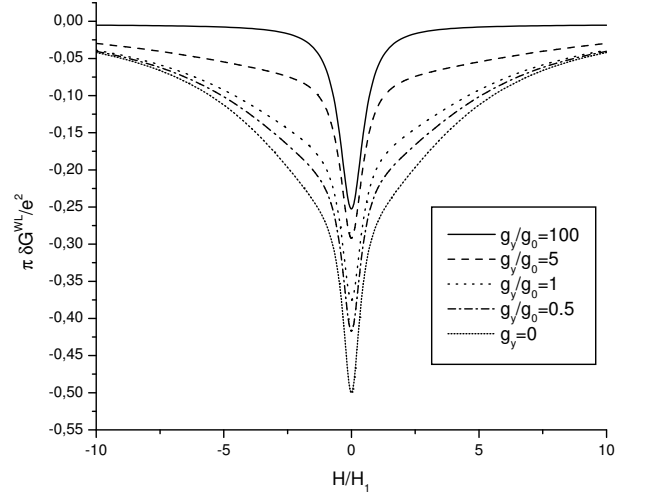}}
\caption{The magnetoconductance of two dots of Fig. \ref{twodotgen}
for $d_1,d_2\gg l_e,$ $d_1/d_2=5$, $g_{ij}=g_0$, $\beta_{ij}=0$, $\beta_y=0$, $\tau_{\varphi 1}=\tau_{\varphi 2}=\infty$.
Here $H_1=1/4\sqrt{\alpha_1\tau_{D1}}$ is the field at which weak localization is
effectively suppressed in the first dot. 
For $g_y=0$ the magnetoconductance is given by superposition of two 
Lorentzians with different widths (decoupled dots), while 
for large $g_y$ only one Lorentzian survives corresponding to the contribution
of a one ``composite dot''.}
\label{cond2dots}
\end{figure}

WL correction to the conductance of the whole structure $\delta
G^{WL}$ is obtained from the general expression for the
conductance determined by Ohm's law:
\begin{eqnarray}
G&=&\big[G_{11}G_{12}(G_{21}+G_{22})+G_{21}G_{22}(G_{11}+G_{12})
\nonumber\\ &&
+\,G_y(G_{12}+G_{22})(G_{11}+G_{21})\big]
\nonumber\\ &&
\big/\,\big[ (G_{11}+G_{12})(G_{21}+G_{22})
\nonumber\\ &&
+\,G_y(G_{11}+G_{12}+G_{21}+G_{22}) \big].
\label{GOhm}
\end{eqnarray}
Substituting $G_{ij}\to G_{ij}+\delta G_{ij}^{WL}$ into this
formula and expanding the result to the first order in $\delta
G_{ij}^{WL}$, we get
\begin{eqnarray}
 \delta G^{WL}=\sum_{i,j=1,2}\frac{\partial G}{\partial  G_{ij}}\delta G_{ij}^{WL}
+\frac{\partial G}{\partial G_{y}}\delta G_{y}^{WL}.
\label{dwl2}
\end{eqnarray}
Combining Eqs. (\ref{Gij})-(\ref{dwl2}) we arrive at the final
result for the WL correction to the conductance of the whole
structure. This general result is rather cumbersome. It is illustrated
in Fig. \ref{cond2dots} for a particular choice of the system parameters.
Below we will specifically consider two important limits.

\begin{figure}
\centerline{\includegraphics[width=7.5cm]{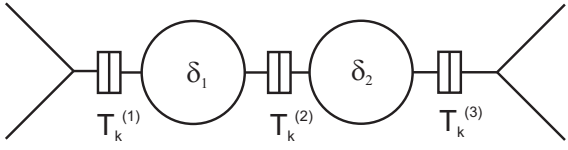}}
\caption{Two quantum dots in series.}
\label{twodot1}
\end{figure}

First we analyze the system of two quantum dots connected in
series, as shown in Fig. \ref{twodot1}, i.e. in the general
structure of Fig. \ref{twodotgen} we set $G_{12}=G_{21}=0,$
$G_{11}=G_1,$ $G_y=G_2$, $G_{22}=G_3$, $\beta_{11}=\beta_1$,
$\beta_y=\beta_2$ and $\beta_{22}=\beta_3$. We also assume $H=0$
and $\tau_\varphi=\infty.$ WL corrections to the barrier
conductances then take the form
\begin{eqnarray}
\delta
G_1^{WL}&=&-\frac{e^2}{\pi}\frac{g_1(g_2+g_3)(1-\beta_1)}{g_1g_2+g_2g_3+g_1g_3},
\nonumber\\
\delta
G_2^{WL}&=&-\frac{e^2}{\pi}\frac{g_2(g_1+g_3)(1-\beta_2)+2g_2^2}{g_1g_2+g_2g_3+g_1g_3},
\nonumber\\
\delta
G_3^{WL}&=&-\frac{e^2}{\pi}\frac{g_3(g_1+g_2)(1-\beta_3)}{g_1g_2+g_2g_3+g_1g_3},
\end{eqnarray}
while Eq. (\ref{GOhm}) reduces to
\begin{eqnarray}
G&=&\frac{G_1G_{2}G_{3}} { G_{1}G_{2} +G_1G_{3}+G_{2}G_3}.
\label{GOhm2}
\end{eqnarray}
WL correction for the whole system then reads
\begin{eqnarray}
\delta G^{WL}&=&-\frac{e^2}{\pi}\frac{g_1g_2^2g_3^2(g_2+g_3)(1-\beta_1)}{(g_1g_2+g_2g_3+g_1g_3)^3}
\nonumber\\ &&
-\,\frac{e^2}{\pi}\frac{g_1^2g_2g_3^2(g_1+g_3)(1-\beta_2)}{(g_1g_2+g_2g_3+g_1g_3)^3}
\nonumber\\ &&
-\, \frac{e^2}{\pi}\frac{g_1^2g_2^2g_3(g_1+g_2)(1-\beta_3)}{(g_1g_2+g_2g_3+g_1g_3)^3}
\nonumber\\ &&
-\, \frac{2e^2}{\pi}\frac{g_1^2g_2^2g_3^2}{(g_1g_2+g_2g_3+g_1g_3)^3}.
\label{2qd}
\end{eqnarray}
In the limit of open quantum dots, i.e. $\beta_{1,2,3}=0$, we
reproduce the result of Ref. \onlinecite{Argaman1}. It is easy to see that
provided the conductance of one of the barriers strongly exceeds
two others, Eq. (\ref{2qd}) reduces to Eq. (\ref{1qd}). If all
three barriers are tunnel junctions, $\beta_{1,2,3} \to 1$, the
first three contributions in Eq. (\ref{2qd}) vanish, and only the
last contribution -- independent of the Fano factors -- survives
in this limit. If, on top of that, one of the tunnel junctions,
e.g. the central one, is less transparent than two others, $g_2
\ll g_1,g_3$, the result acquires a particularly simple (non-Lorentzian) form
\begin{eqnarray}
\delta G^{WL}=
- \frac{2e^2}{\pi}\frac{g_2^2}{\left(g_1+\gamma_1\right)\left(g_3+\gamma_2\right)},
\end{eqnarray}
with $\gamma_{1,2}$ defined in Eq. (\ref{gamma}).
Note that  $\delta G^{WL}\propto g_2^2$, i.e. this result is dominated by the 
second order tunneling processes across the second barrier.

\begin{figure}
\centerline{\includegraphics[width=3cm]{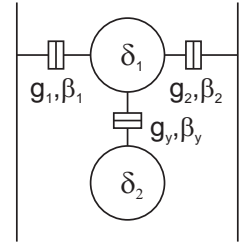}} \caption{The
system of two connected quantum dots only one of which is in turn
connected to the leads.} \label{twodot11}
\end{figure}

Our second example is the system depicted in Fig. \ref{twodot11}
which corresponds to the following choice of parameters in Fig.
\ref{twodotgen}: $G_{11}=G_1,$ $G_{12}=G_2,$ $G_{21}=G_{22}=0,$
$\beta_{11}=\beta_1$ and $\beta_{12}=\beta_2$. In addition, we
assume that electrons are subject to dephasing only in the second quantum dot, i.e. 
$\tau_{\varphi 1}=\infty$ while $\tau_{\varphi 2}$ is finite. 
This setup allows one to analyze the so-called dephasing by voltage probes
\cite{But,BB1}. We obtain
\begin{eqnarray}
C_{11}=\frac{4\pi}{\delta_1}\frac{g_y+4\pi/\delta_2\tau_{\varphi 2}}
{(g_1+g_2)g_y+4\pi(g_1+g_2+g_y)/\delta_2\tau_{\varphi 2}}.
\end{eqnarray}
In the limit $\tau_{\varphi 2} \to \infty$ this result reduces to
\begin{eqnarray}
C_{11}=\frac{4\pi}{\delta_1}\frac{1}{g_1+g_2}
\end{eqnarray}
and we again arrive at Eq. (\ref{1qd}), i.e. the second quantum
dot attached to the first one does not affect the expression for
WL correction. In the opposite limit of short decoherence times,
$\tau_{\varphi 2}\to 0$,
we find
\begin{eqnarray}
C_{11}=\frac{4\pi}{\delta_1}\frac{1}{g_1+g_2+g_y}
\end{eqnarray}
and arrive at the WL correction\cite{BB1}
\begin{eqnarray}
\delta G^{WL}=-\frac{e^2}{\pi}\frac{g_1g_2^2(1-\beta_1)+g_1^2g_2(1-\beta_2)}{(g_1+g_2)^3
\left(1+{\tau_D}/{\tau_\varphi^{\rm eff}}\right)},
\label{WLdef}
\end{eqnarray}
where
\begin{eqnarray}
\frac{1}{\tau_\varphi^{\rm eff}}=\frac{g_y}{g_1+g_2}\frac{1}{\tau_D}
\end{eqnarray}
is the electron decoherence rate induced in the first
quantum dot due to coupling to the second one acting as an effective
voltage probe.

\subsection{1D array of identical quantum dots}

Let us now turn to 1d arrays of quantum dots depicted in
Fig. \ref{array1}. For simplicity, we will assume that our array
consists of $N-1$ identical quantum dots with the same level
spacing $\delta_n\equiv \delta_d$ and of $N$ identical barriers
with the same dimensionless conductance $g_n\equiv g$ and the same
Fano factor $\beta_n\equiv \beta$. We will also assume that the
quantum dots have the same shape and size so that 
$\tau_{Hn}\equiv \tau_H$ and $\tau_{\varphi n}\equiv\tau_\varphi$.
For this system the Cooperon can also
be found exactly. The result reads
\begin{eqnarray}
C_{nm}(\omega)=\frac{2}{N}\sum_{q=1}^{N-1}\frac{\sin\frac{\pi qn}{N}\sin\frac{\pi qm}{N}}
{-i\omega+\frac{1}{\tau_H}+\frac{1}{\tau_\varphi}+\frac{1-\cos\frac{\pi q}{N}}{\tau_D}}.
\end{eqnarray}
Here $\tau_D=2\pi /g\delta_d$ and $\tau_H=1/16\alpha H^2.$

The WL correction then takes the form
\begin{eqnarray}
\delta G^{WL}=-\frac{e^2g\delta_d}{2\pi^2N^2}\sum_{q=1}^{N-1}
\frac{\beta\cos\frac{\pi q}{N}+1-\beta}
{\frac{1}{\tau_H}+\frac{1}{\tau_\varphi}+\frac{1-\cos\frac{\pi
q}{N}}{\tau_D}}. \label{WLN}
\end{eqnarray}

The sum over $q$ can be handled exactly and yields
\begin{eqnarray}
&& \delta G^{WL}=-\frac{e^2}{\pi N^2}
\bigg[
\left(N\frac{1+u^{2N}}{1-u^{2N}}-\frac{1+u^2}{1-u^2}\right)
\nonumber\\ &&\times\,
\frac{\beta(1+u^2)+2(1-\beta)u}{1-u^2} -(N-1)\beta
\bigg],
\label{magresN}
\end{eqnarray}
where
\begin{eqnarray}
u=1+\frac{\tau_D}{\tau_H}+\frac{\tau_D}{\tau_\varphi}
-\sqrt{\left(1+\frac{\tau_D}{\tau_H}+\frac{\tau_D}{\tau_\varphi}\right)^2-1}.
\label{uuu}
\end{eqnarray} In the tunneling limit $\beta=1$ and for
$\tau_\varphi\to\infty$ our result defined in Eqs. (\ref{magresN})-(\ref{uuu}) 
becomes similar -- though not exactly
identical -- to the corresponding result \cite{CN}. 

If $\tau_\varphi$ is long enough, namely $
{1}/{\tau_\varphi}\lesssim E_{\rm Th}, $ where $E_{\rm
Th}=\pi^2/2N^2\tau_D$ is the Thouless energy of the whole array,
in Eqs. (\ref{WLN})-(\ref{magresN}) it is sufficient to set
$\tau_\varphi=\infty$. In this case the magnetic field $H$
significantly suppresses WL correction provided $1/\tau_H\gtrsim
E_{\rm Th}$ or, equivalently, if
\begin{eqnarray}
H\gtrsim H_N,\;\; H_N=\frac{1}{8N}\sqrt{\frac{\pi
g\delta_d}{\alpha}}. \label{HN}
\end{eqnarray}

\begin{figure}
\centerline{\includegraphics[width=8cm]{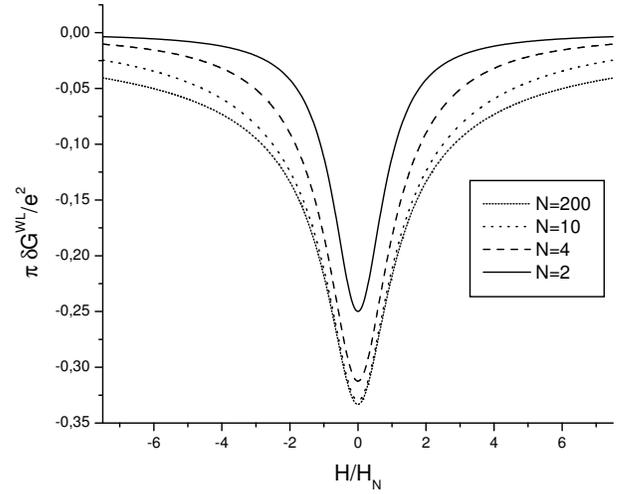}}
\caption{Magnetoconductance of a 1d array of $N-1$ identical open
($\beta=0$) quantum dots in the absence of interactions
($\tau_\varphi\to\infty$). The field $H_N$ is defined in Eq.
(\ref{HN}). } \label{twodot}
\end{figure}

In the opposite limit $1/\tau_\varphi\gtrsim E_{\rm Th}$
we find
\begin{eqnarray}
\delta G^{WL}=-\frac{e^2}{\pi N}\left[
\frac{\beta\left(1+\frac{\tau_D}{\tau_H}+\frac{\tau_D}{\tau_\varphi}\right)+1-\beta}
{\sqrt{\left(1+\frac{\tau_D}{\tau_H}+\frac{\tau_D}{\tau_\varphi}\right)^2-1}}
-\beta\right].
\end{eqnarray}
In particular, in the diffusive limit
$\tau_H,\tau_\varphi\gg\tau_D$ we get
\begin{eqnarray}
\delta G^{WL}=-\frac{e^2}{\pi
Nd}\sqrt{\frac{D\tau_H\tau_\varphi}{\tau_H+\tau_\varphi}},
\label{diffmet}
\end{eqnarray}
where we introduced the diffusion coefficient
\begin{eqnarray}
D=d^2/2\tau_D.
\label{D}
\end{eqnarray}
Eq. (\ref{diffmet}) coincides with the standard result for
quasi-1d diffusive metallic wire. Note, however, that the values
of $\tau_H$ within our model may differ from those for a metallic
wire. The ratio of the former to the latter is $\tau_H^{\rm
qd}/\tau_H^{\rm met}\sim \tau_{\rm fl}/\tau_D,$ where $\tau_{\rm
fl}\sim d/v_F$ is the flight time through the quantum dot. Since
typically $\tau_{\rm fl}<\tau_D$ we conclude that for the same
value of $D$ the magnetic field dephases electrons stronger in the
case of an array of quantum dots.

For a single quantum dot ($N=2$) Eq. (\ref{magresN})
reduces to
\begin{eqnarray}
\delta G^{WL}=-\frac{e^2(1-\beta)}{4\pi}
\frac{1}{\left(1+\frac{\tau_D}{\tau_H}+\frac{\tau_D}{\tau_\varphi}\right)}
\end{eqnarray}
in agreement with Eq. (\ref{magres1qd}).

For two identical quantum dots in series we obtain
\begin{eqnarray}
\delta G^{WL}=-\frac{e^2}{9\pi}\bigg[
\frac{2-\beta}{1+\frac{2\tau_D}{\tau_H}+\frac{2\tau_D}{\tau_\varphi}}
+\frac{\frac{2}{3}-\beta}{1+\frac{2\tau_D}{3\tau_H}+\frac{2\tau_D}{3\tau_\varphi}}
\bigg],
\end{eqnarray}
i.e. the magnetoconductance is just the sum of two Lorentzians in
this case.

Finally, in the absence of any interactions ($\tau_\varphi=\infty$) and
at $H=0$ we obtain
\begin{eqnarray}
\delta G^{WL}=-\frac{e^2}{\pi}\left[\frac{1}{3}-\frac{\beta}{N}
+\frac{1}{N^2}\left(\beta-\frac{1}{3}\right)\right].
\label{GWLarray}
\end{eqnarray}
In the limit $N\to \infty$ this result reduces to the standard one
for a long quasi-1d diffusive wire \cite{Mello} while for any finite $N$ we
reproduce the results for tunnel barriers \cite{CN} ($\beta \to 1$) and 
open quantum dots \cite{Argaman2} ($\beta \to 0$).

\section{Generalization to 2d arrays}

\begin{figure}
\centerline{\includegraphics[width=7cm]{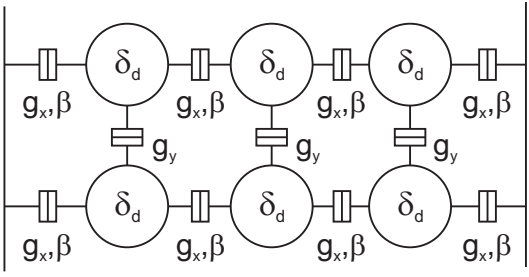}} \caption{2d
array of identical quantum dots. Here the number of barriers in
the $x$-direction is chosen to be $N=4$, and the number of quantum
dots in the $y-$direction is $M=2$. The barriers are characterized by
the dimensionless conductances $g_x$ and $g_y$ and the Fano factor
$\beta$.}
\end{figure}

Until now our analysis was only focused on structures with several
quantum dots and 1d arrays. Generalization to the case of 
2d and 3d systems is straightforward. Below we analyze an
important case of 2d arrays.

Consider an array consisting of $N-1\times M$ quantum dots. For
simplicity, here we will only deal with the case of identical
quantum dots, see Fig. 7. The WL correction to the conductance of
this array reads
\begin{eqnarray}
\delta G^{WL}=\frac{1}{N^2}\sum_{n=1}^N\sum_{m=1}^M \delta G^{WL}_{nm},
\label{G2d}
\end{eqnarray}
where, similarly to Eq. (\ref{GWL}),
\begin{eqnarray}
\delta G^{WL}_{nm}&=&-\frac{e^2 g_x\delta_d}{4\pi^2 }\int_{0}^\infty dt
\big\{ \beta\big[ C_{n-1,n;mm}(t)
\nonumber\\ &&
+\, C_{n,n-1;mm}(t)\big]
+(1-\beta)\big[ C_{nn;mm}(t)
\nonumber\\ &&
+\,C_{n-1,n-1;mm}(t)\big]\big\}
\label{GWLnm}
\end{eqnarray}
defines the WL correction for the barrier with ``coordinates'' $n,m$. In order
to find the Cooperons  $C_{nn';mm'}(t)$ one needs to solve the equation
\begin{eqnarray}
&& \left(\frac{\partial}{\partial t}+\frac{1}{\tau_H}+\frac{1}{\tau_\varphi}\right)C_{nn';mm'}
+\,\frac{\delta_d}{4\pi}\big[(2g_x+2g_y)
\nonumber\\ &&
\times\,C_{nn';mm'}-g_xC_{n-1,n';mm'}-g_{x}C_{n+1,n';mm'}
\nonumber\\ &&
-\,g_yC_{n,n'-1;mm'}-g_{y}C_{n,n'+1;mm'}
\big]
=\delta_{nn'}\delta_{mm'}\delta(t).\hspace{0.65cm}
\label{diff2d}
\end{eqnarray}
which is directly analogous to Eq. (\ref{diff}). In the zero frequency limit
the solution of this equation with appropriate boundary conditions reads
\begin{eqnarray}
&& C_{{nn'};{mm'}}(\omega\to 0)=\frac{2}{MN}\sum_{q_x=1}^{N-1}\sum_{q_y=0}^{M-1}
\nonumber\\ &&
\frac{
\sin\frac{\pi q_xn'}{N}\left[\cos\frac{\pi q_ym'}{M}+\cos\frac{\pi q_y(m'-1)}{M}\right]}
{\frac{1}{\tau_H}+\frac{1}{\tau_\varphi}+\frac{g_x\delta_d}{2\pi}\left(1-\cos\frac{\pi q_x}{N}\right)
+\frac{g_y\delta_d}{2\pi}\left(1-\cos\frac{\pi q_y}{M}\right)}
\nonumber\\ &&\times\,
\frac{\sin\frac{\pi q_xn}{N}\left[\cos\frac{\pi q_ym}{M}+\cos\frac{\pi q_y(m-1)}{M}\right]}{\delta_{0,q_y}+1+\cos\frac{\pi q_y}{M}}.
\label{C2d}
\end{eqnarray}

Combining Eqs. (\ref{G2d}), (\ref{GWLnm})  and (\ref{C2d}), we obtain
\begin{eqnarray}
 &&\delta G^{WL}= -\frac{e^2 g_x\delta_d}{2\pi^2 N^2} \sum_{q_x=1}^{N-1}\sum_{q_y=0}^{M-1}
\nonumber\\ &&
\frac{\beta\cos\frac{\pi q_x}{N}+1-\beta}
{\frac{1}{\tau_\varphi}+\frac{1}{\tau_H}+\frac{g_x\delta_d}{2\pi}\left(1-\cos\frac{\pi q_x}{N}\right)
+\frac{g_y\delta_d}{2\pi}\left(1-\cos\frac{\pi q_y}{M}\right)}.\;\;\;\;\;\;\;
\label{WL2d}
\end{eqnarray}
This result is valid provided an external magnetic field $H$
dephases electrons predominantly inside
quantum dots. This is the case provided the field is not too
low \cite{blanter,Varlamov}, $H\gtrsim H^*$, where
$H^*=\pi c g\delta\tau_{\rm fl}/ed^2.$ At lower fields, $H<H^*$,
one can apply the standard  theory \cite{AA} developed for homogeneous
metals, in which case $1/\tau_H^{\rm met}=4eDH/c$, where
$D$ is now defined in Eq. (\ref{D}). Substituting the value $H^*$ into
this expression we get $1/\tau_H^*\sim \tau_{\rm fl}/\tau_D^2.$
Comparing this energy scale with $1/\tau_\varphi$ and $E_{\rm Th}$
we immediately arrive at the condition
\begin{eqnarray}
\max\{1/\tau_\varphi,E_{\rm Th}\}\gg g^2\delta^2\tau_{\rm fl},
\end{eqnarray}
for which Eq. (\ref{WL2d}) is applicable at all values of $H$.

Turning to concrete examples we first consider the simplest case
with $N=2, M=2$, which is a symmetric version of the system
of Fig. \ref{twodotgen}. Eq. (\ref{WL2d}) then yields
\begin{eqnarray}
 \delta G^{WL}&=& -\frac{e^2 g_x\delta_d}{8\pi^2}
\frac{1-\beta}
{\frac{1}{\tau_\varphi}+\frac{1}{\tau_H}+\frac{g_x\delta_d}{2\pi}}
\nonumber\\ && -\,\frac{e^2 g_x\delta_d}{8\pi^2} \frac{1-\beta}
{\frac{1}{\tau_\varphi}+\frac{1}{\tau_H}+\frac{g_x\delta_d}{2\pi}
+\frac{g_y\delta_d}{2\pi}}.
\end{eqnarray}
For $H=0$ and $\tau_\varphi\to\infty$ this expression reduces to
\begin{eqnarray}
 \delta G^{WL}= -\frac{e^2 (1-\beta)}{4\pi}\left(1+\frac{g_x}{g_x+g_y}\right).
\end{eqnarray}

Next we consider an extended isotropic ($g_x=g_y$) 2d array  and stick to the diffusive
regime
$$
E_{\rm Th}\lesssim \frac{1}{\tau_\varphi}+\frac{1}{\tau_H}
\lesssim \frac{1}{\tau_D}.
$$
In this case we find
\begin{eqnarray}
\delta G^{WL}=-\frac{e^2}{2\pi^2}\frac{M}{N}
\left[\ln\frac{\tau_H\tau_\varphi}{\tau_D(\tau_H+\tau_\varphi)}+2.773-\pi\beta\right].
\end{eqnarray}
The leading term in this equation matches
with the standard WL correction for a 2d diffusive metallic film in the
parallel magnetic field \cite{AA}.

Let us briefly discuss an effect of anisotropy. In the limit of
small $g_y\ll g_x/N^2 $ the system reduces to a set of $M$
essentially independent 1d arrays and, hence, $ \delta
G^{WL}=M\delta G^{WL}_{1d}, $ where $\delta G^{WL}_{1d}$ is
defined in Eqs. (\ref{WLN},\ref{magresN}). In the opposite limit
of large $g_y\gg g_xM^2$ electron diffusion in the direction
perpendicular to the current becomes fast, and one can treat the
system as a 1d array of $N-1$ composite quantum dots each of them
consisting of $M$ original dots. In this limit we get $ \delta
G^{WL}=\delta G^{WL}_{1d}. $

Finally, let us note that our Eq. (\ref{WL2d}) also
allows to reproduce recent results \cite{Varlamov} for WL correction
to the conductivity of bulk granular metals. In
order to handle this limit, in Eq. (\ref{WL2d}) one should formally set 
$M,N\to \infty$ (which yields $\delta G^{WL}\propto M/N$ and
 allows to define the conductivity) and then put $\beta=1$ and
$g_x=g_y$.

\section{Summary}

In this paper we have developed a theoretical approach 
based on a combination
of the scattering matrix formalism with the non-linear $\sigma$-model
technique. This approach allows to analyze weak localization effects for an arbitrary system of quantum dots connected via barriers
with arbitrary distribution of channel transmissions. This general model can
be used to describe virtually any type of disordered conductors. 
Employing our approach we have evaluated WL corrections to the system
conductance in a number of important physical situations, 
e.g., for the case of two quantum dots connected to each other
and to external leads in an arbitrary way (Sec. V B), as well as 
for 1d (Sec. V C) and 2d (Sec. VI) arrays of identical quantum dots. In a
number of specific limits our general results reduce to those derived earlier
by means of other approaches.

The results obtained here remain valid either in the absence of interactions or
provided the interaction effects on weak localization are taken into account
within a phenomenological scheme which amounts to introducing electron
decoherence time $\tau_\varphi$ as an additional parameter. The method
proposed here also serves as a good starting point for a more general
and systematic analysis of electron-electron interaction effects. This
analysis will be worked out in our forthcoming publications.

This work is part of the EU Framework Programme
NMP4-CT-2003-505457 ULTRA-1D "Experimental and theoretical investigation of
electron transport in ultra-narrow 1-dimensional nanostructures".

\appendix

\section{Relation between Green function $\check {\cal G}$ and $\check Q$.}

Here we will closely follow the method proposed by
Nazarov\cite{Nazarov}. Let us select one of the barriers in our
array and denote (coordinate-independent) $Q-$fields in the left
(right) dot with respect to this barrier as $\check Q_L$ ($\check
Q_R$). Provided $\check Q_{L,R}$ are slow functions of time, in
the barrier vicinity one can neglect the term $i\partial/\partial
t$. In addition, one can linearize the electron spectrum in the
vicinity of the Fermi energy and replace $\nabla^2/2m \to \pm
iv_m\partial/\partial x,$ where $v_n=p_n/m$ is the electron
velocity in a given channel. As a result, for the left dot one
gets
\begin{eqnarray}
\left(i\alpha v_n\frac{\partial}{\partial x}+\frac{i}{2\tau_e}\check Q_L\right)
\check{\cal G}_{nm}^{\alpha\beta}=\delta(x-x')\delta_{nm}\delta_{\alpha\beta}.
\end{eqnarray}
Defining the diagonal matrix $\hat v=v_n\delta_{nm}$, and making
use of the normalization condition (\ref{norm}), we can write the
solution in the form
\begin{eqnarray}
&& \check{\cal G}^{\alpha\beta}_L(x,x')=
\nonumber\\ &&
\frac{1}{4}
\left[e^{\hat v^{-1}x/2\tau_e}(1-\alpha\check Q_L)
+e^{-\hat v^{-1}x/2\tau_e}(1+\alpha\check Q_L) \right]
\nonumber\\ &&\times\,
\left[-i\alpha \delta_{\alpha\beta}\hat v^{-1}\theta(x-x') + \check R_L^{\alpha\beta}\right]
\nonumber\\ &&\times\,
\left[e^{\hat v^{-1}x'/2\tau_e}(1+\alpha\check Q_L)
+e^{-\hat v^{-1}x'/2\tau_e}(1-\alpha\check Q_L) \right].\hspace{0.7cm}
\end{eqnarray}
Here $\check R_L^{\alpha\beta}$ is an arbitrary operator.
Requiring $\check{\cal G}^{\alpha\beta}_L$ not to grow
exponentially far from the barrier we arrive at the following
constraints:
\begin{eqnarray}
\left(
\begin{array}{cc}
1+\check Q_L & 0 \\
0 & 1-\check Q_L
\end{array}
\right)
\left(
\begin{array}{cc}
\check R_L^{++} & \check R_L^{+-} \\
\check R_L^{-+} & \check R_L^{--}
\end{array}
\right)=0,
\nonumber\\
\left(
\begin{array}{cc}
\check R_L^{++} & \check R_L^{+-} \\
\check R_L^{-+} & \check R_L^{--}
\end{array}
\right)
\left(
\begin{array}{cc}
1-\check Q_L & 0 \\
0 & 1+\check Q_L
\end{array}
\right)=0.
\label{constL}
\end{eqnarray}
Similarly, for the right dot we obtain
\begin{eqnarray}
\left(
\begin{array}{cc}
1-\check Q_R & 0 \\
0 & 1+\check Q_R
\end{array}
\right)
\left(
\begin{array}{cc}
\check R_R^{++} & \check R_R^{+-} \\
\check R_R^{-+} & \check R_R^{--}
\end{array}
\right)=0,
\nonumber\\
\left(
\begin{array}{cc}
\check R_R^{++} & \check R_R^{+-} \\
\check R_R^{-+} & \check R_R^{--}
\end{array}
\right)
\left(
\begin{array}{cc}
1+\check Q_R & 0 \\
0 & 1-\check Q_R
\end{array}
\right)=0.
\label{constR}
\end{eqnarray}
Note  that the elastic mean free time $\tau_e$ drops out of Eqs.
(\ref{constL},\ref{constR}), thus indicating a very general
nature of these constraints.
The Green functions on the left and right barrier sides are
related to each other by the $S-$matrix
\begin{eqnarray}
\hat S=\left(
\begin{array}{cc}
\hat r & \hat t' \\
\hat t & \hat r'
\end{array}
\right)
\end{eqnarray}
of this barrier. This relation has the form
\begin{eqnarray}
 \sqrt{\hat v} \check {\cal G}^{\alpha\beta}_R \sqrt{\hat v}=
\hat M \sqrt{\hat v} \check {\cal G}^{\alpha\beta}_L \sqrt{\hat v}\hat M^\dagger,
\label{transfer}
\end{eqnarray}
\begin{eqnarray}
\hat M=\left(
\begin{array}{cc}
\hat t-\hat r'\hat t^{\prime -1}\hat r & \hat r'\hat t^{\prime -1} \\
-\hat t^{\prime -1}\hat r & \hat t^{\prime -1}
\end{array}
\right)
\end{eqnarray}
being the transfer matrix which satisfies $\hat
M\sigma_z\hat M^\dagger=\sigma_z.$

Eqs. (\ref{constL}), (\ref{constR}) and (\ref{transfer}) for $\check
R_{L,R}^{\alpha\beta}$ can be resolved making use of the fact that
in the barrier vicinity, i.e. for $|x|,|x'|\ll v_n\tau_e$, the
Green function takes the form
\begin{eqnarray}
&& \check {\cal G}^{\alpha\beta}_{L,R}=
\left(
\begin{array}{cc}
\check R_{L,R}^{++} & \check R_{L,R}^{+-} \\
\check R_{L,R}^{-+} & \check R_{L,R}^{--}
\end{array}
\right)
\nonumber\\ &&
+\,
\frac{i\hat v^{-1}}{2}
\left(
\begin{array}{c}
-{\rm sign} (x_1-x_2)-\check Q_{L,R} \hspace{1cm} 0 \\
0 \hspace{1cm} {\rm sign}(x_1-x_2) - \check Q_{L,R}
\end{array}
\right). \label{GLR}
\end{eqnarray}
The operators $\check R_{L,R}^{\alpha\beta}$ turn out to be
diagonal in the channel indices in the basis for which the
matrices $\hat t$ and $\hat r$ are diagonal as well. Defining the
channel transmission values $T_n=|t_n|^2$, we get
\begin{eqnarray}
&& \check R_{L,nm}^{\alpha\beta}=-\frac{i}{v_n}\delta_{nm}
\left[1+\frac{T_n}{4}\big(\{\check Q_L,\check Q_R\}-2\big)\right]^{-1}
\nonumber\\ &&\times\,
\left(
\begin{array}{c}
\frac{T_n\big(\check Q_R+[\check Q_R,\check Q_L]-\check Q_L\check Q_R\check Q_L\big)}{8}
\hspace{0.3cm}
-\frac{r^*_n(1-\check Q_L)}{2} \\
\frac{r_n(1+\check Q_L)}{2}
\hspace{0.3cm}
\frac{T_n\big(\check Q_R-[\check Q_R,\check Q_L]-\check Q_L\check Q_R\check Q_L\big)}{8}
\end{array}
\right),\hspace{1cm}
\label{RL}
\end{eqnarray}
\begin{eqnarray}
&& \check R_{R,nm}^{\alpha\beta}=-\frac{i}{v_n}\delta_{nm}
\left[1+\frac{T_n}{4}\big(\{\check Q_L,\check Q_R\}-2\big)\right]^{-1}
\nonumber\\ &&\times\,
\left(
\begin{array}{c}
\frac{T_n\big(\check Q_L+[\check Q_R,\check Q_L]-\check Q_R\check Q_L\check Q_R\big)}{8}
\hspace{0.3cm}
\frac{r'_n(1+\check Q_R)}{2} \\
-\frac{r^{\prime *}_n(1-\check Q_R)}{2}
\hspace{0.3cm}
\frac{T_n\big(\check Q_L-[\check Q_R,\check Q_L]-\check Q_R\check Q_L\check Q_R\big)}{8}
\end{array}
\right).\hspace{1cm}
\label{RR}
\end{eqnarray}

In order to find WL correction (\ref{GWL3}) it is sufficient
to determine the Green function  (\ref{RL}) only in the left  dot:
\begin{eqnarray}
\check R^{++}_{L,nm}&=&\frac{T_n\delta_{nm}}{2v_n}
\left(
\begin{array}{cccc}
0 & 2\Delta u_{1} & 0 & 0 \\
0 & 0 & 0 & 0 \\
0 & g^K\Delta u_{1} & 0 & 0 \\
-\Delta v_{2}g^K & 0 & 2\Delta v_{2} & 0
\end{array}
\right),
\nonumber\\
\check R^{--}_{L,nm}&=&\frac{T_n\delta_{nm}}{2v_n}
\left(
\begin{array}{cccc}
0 & 0 & 0 & 0 \\
-2\Delta u_{2} & 0 & 0 & 0 \\
0 & \Delta v_{1}g^K & 0 & -2\Delta v_1 \\
-g^K\Delta u_{2} & 0 & 0 & 0
\end{array}
\right),
\nonumber\\
\check R^{+-}_{L,nm}&=&\frac{ir^*_n\delta_{nm}}{2v_n}\check A,\;\;\;
\check R^{-+}_{L,nm}=\frac{ir_n\delta_{nm}}{2v_n}\check A,
\label{Rpm}
\end{eqnarray}
where $\Delta u_{1,2}=u_{1,2R}-u_{1,2L}$,
$\Delta v_{1,2}=v_{1,2R}-v_{1,2L}$ and
\begin{eqnarray}
\check A=i\left(
\begin{array}{cccc}
-2i & -2 u_{1L} & 0 & 0 \\
2u_{2L} & 0 & 0 & 0 \\
-ig^K & -v_{1L}g^K-g^Ku_{1L} & 0 & 2v_{1L} \\
v_{2L}g^K+g^Ku_{2L} & ig^K & -2 v_{2L} & -2i
\end{array}
\right).
\nonumber
\end{eqnarray}
Here $R_{L,nm}^{\alpha,\beta}$
has been expanded to the first order in $u_{1,2}$ and $v_{1,2}$.
From Eqs. (\ref{GLR}) and (\ref{Rpm}) we find
\begin{eqnarray}
 & {\cal G}_{L,nn;34}^{++}=-\frac{v_{1L}}{v_n}, &
{\cal G}_{L,nn;34}^{--}=\frac{-v_{1L}-T_n\Delta v_1}{v_n},
\nonumber\\
& {\cal G}_{L,nn;34}^{+-}=-r^*_n\frac{v_{1L}}{v_n},&
{\cal G}_{L,nn;34}^{-+}=-r_n\frac{v_{1L}}{v_n};
\nonumber\\
& {\cal G}_{L,nn;43}^{++}=\frac{v_{2L}+T_n\Delta v_{2}}{v_n},&
{\cal G}_{L,nn;43}^{--}=\frac{v_{2L}}{v_n},
\nonumber\\
& {\cal G}_{L,nn;43}^{+-}=r^*_n\frac{v_{2L}}{v_n}, &
{\cal G}_{L,nn;43}^{-+}=r_n\frac{v_{2L}}{v_n}.
\end{eqnarray}
Substituting these expressions into Eq. (\ref{GWL4}), after some
transformations we arrive at Eq. (\ref{GWL5}).

\end{document}